\documentclass{article}
\usepackage{ijcai19}

\usepackage{times}
\usepackage{soul}
\usepackage{url}
\usepackage[utf8]{inputenc}
\usepackage[small]{caption}
\usepackage{graphicx}
\usepackage{subfigure}
\usepackage{amssymb}
\usepackage{amsmath}
\usepackage{booktabs}
\usepackage{algorithmic}
\usepackage[ruled,vlined]{algorithm2e}
\usepackage{enumitem}

\setlist[itemize]{itemsep=-0.25ex,leftmargin=2.5ex}
\setlist[enumerate]{itemsep=-0.25ex,leftmargin=2.5ex}

\title{Diversity-Promoting Deep Reinforcement Learning for Interactive Recommendation}

\author{
    Yong Liu$^\dag$, Yinan Zhang$^\ddag$, Qiong Wu$^\dag$, Chunyan Miao$^\dag$, Lizhen Cui$^\ddag$,\\Binqiang Zhao$^\sharp$, Yin Zhao$^\sharp$, Lu Guan$^\sharp$\\
    \affiliations
    $^\dag$Nanyang Technological University, Singapore\\
    $^\ddag$Shandong University, China\\
    $^\sharp$Alibaba Group, China
   \emails
 \{stephenliu,wu.qiong, ascymiao\}@ntu.edu.sg, zhangynnancy@outlook.com, clz@sdu.edu.cn,\\ \{binqiang.zhao, zhaoyin.zy, guanlu.gl\}@alibaba-inc.com
}

\begin{document}

\maketitle

\begin{abstract}
Interactive recommendation that models the explicit interactions between users and the recommender system has attracted a lot of research attentions in recent years. Most previous interactive recommendation systems only focus on optimizing recommendation accuracy while overlooking other important aspects of recommendation quality, such as the diversity of recommendation results. In this paper, we propose a novel recommendation model, named \underline{D}iversity-promoting \underline{D}eep \underline{R}einforcement \underline{L}earning (D$^2$RL), which encourages the diversity of recommendation results in interaction recommendations. More specifically, we adopt a Determinantal Point Process (DPP) model to generate diverse, while relevant item recommendations. A personalized DPP kernel matrix is maintained for each user, which is constructed from two parts: a fixed similarity matrix capturing item-item similarity, and the relevance of items dynamically learnt through an actor-critic reinforcement learning framework. We performed extensive offline experiments as well as simulated online experiments with real world datasets to demonstrate the effectiveness of the proposed model.
\end{abstract}

\section{Introduction}
Recommender systems have been widely used in various application scenarios. In practice, existing recommender systems are mainly developed based on collaborative filtering methods, e.g., matrix factorization~\cite{shi2014collaborative}. Recently, with the rapid development of deep learning techniques, e.g., Convolutional Neural Networks (CNN) and Recurrent Neural Networks (RNN), a lot of deep learning based recommender systems have been proposed~\cite{zhang2017deep}. Particularly, most recommender systems are developed in an off-line manner, by learning users' preferences on items from the historical interaction data between users and items. These methods, however, usually ignore the interactions between users and the recommender system. Therefore, they cannot adopt the users' immediate and long-term feedback on the recommendation results to improve the performance of the recommender system.

From a different perspective, the interactive recommender system~\cite{steck2015interactive} explicitly models the interactions between users and the recommender system. In this scenario, the recommender system is treated as an agent, which generates recommendations to users according to some learnt policy. Based on users' feedback on the recommendation results, the agent consequentially update its recommendation policy, thus to improve users' satisfaction in the long run. Moreover, in the interaction process, the interactive recommender system can also explore users' potential preferences on items, which may not be revealed by users' historical behaviour data.

Some of the existing interactive recommender systems are developed based on the contextual bandit framework, e.g., LinUCB~\cite{li2010contextual} and Thompson sampling~\cite{kawale2015efficient}. Recently, some research works build interactive recommender systems using the deep reinforcement learning framework, considering the state transitions of the environment (i.e., users)~\cite{zheng2018drn,zhao2018deep,zhao2018recommendations,chen2018large}. These recommendation methods mainly focus on optimizing the recommendation accuracy. However, they do not enforce the diversity on the recommendation results. Therefore, the items in a recommendation list generated by these methods may usually be very similar with each other. This often results in the reduction of user experiences and user satisfaction.

Differing from existing interactive recommendation methods, this paper proposes a novel recommendation model, namely \underline{D}iversity-promoting \underline{D}eep \underline{R}einforcement \underline{L}earning (D$^2$RL), for diversified interactive recommendation. More specifically, to generate diverse while relevant item recommendations, we do sequential sampling from a Determinantal Point Process (DPP) model, which is an elegant probabilistic model for set diversity~\cite{kulesza2012determinantal}. For each user, a personalized DPP kernel matrix is dynamically built from two parts: a fixed similarity matrix capturing item-item similarity, and the relevance of items learnt through deep reinforcement learning. To dynamically learn the relevance of items, we adopt the actor-critic deep reinforcement learning framework to explore and exploit the user's preferences on items. In D$^2$RL, the actor network generates an action vector, i.e., the parameter of the DPP kernel matrix, which represents the user's dynamic personal preferences. The critic network evaluates the quality of the learnt recommendation policy, considering both current and future rewards for the recommended items. To demonstrate the effectiveness of the proposed D$^2$RL model, we have conducted extensive experiments on real world datasets, as well as simulated online experiments. The experimental results indicate that D$^2$RL has superior performance under different settings.

To summarize, the major contributions of this work are as follows:
\begin{itemize}
  \item We propose a novel diversity-promoting recommendation model, i.e., D$^2$RL, which is able to generate diverse while relevant recommendations in interactive recommendation.
  \item In order to test the online performance of D$^2$RL, we propose a novel online simulator which can simulate user behaviours by considering both user preferences and the diversity of user behaviors. 
  \item Extensive experiments with both offline and online settings demonstrate the superior performance of D$^2$RL in comparison with several state-of-the-art methods.
\end{itemize}

\section{Related Work}
Contextual bandit has often been used for building interactive recommender systems. For instance, Li et al. proposed a contextual bandit algorithm named LinUCB to sequentially recommend articles to users based on the contextual information of users and articles~\cite{li2010contextual}.
Kawale et al. proposed to integrate Thompson sampling with online Bayesian probabilistic matrix factorization for interactive recommendation~\cite{kawale2015efficient}. Tang et al. developed a parameter-free bandit approach which applied online bootstrap to learn the recommendation model in an online manner~\cite{tang2015personalized}. Moreover, in~\cite{wang2017factorization}, a factorization-based bandit approach was proposed to solve the online interactive recommendation problem. Another group of works apply deep reinforcement learning for solving the interactive recommendation problem. For example, Zheng et al. proposed a deep reinforcement learning framework for news recommendation~\cite{zheng2018drn}. They utilized deep Q-learning to explicitly model the future rewards of recommendation results and also developed an effective exploration strategy to improve the recommendation accuracy. Zhao et al. proposed a deep recommendation framework (i.e., DEERS) that leveraged reinforcement learning to consider both users' positive and negative feedbacks for recommendation~\cite{zhao2018recommendations}. Moreover, they also developed a page-wise recommendation model (i.e., DeepPage) which employed deep reinforcement learning to optimize the display of items on a webpage~\cite{zhao2018deep}. In a recent work~\cite{chen2018large}, Chen et al. proposed a tree structured policy gradient method to mitigate the large action space problem of deep reinforcement learning based recommendation methods.

Promoting the diversity of recommendation results has received increasing research attentions~\cite{kunaver2017diversity}. The maximal marginal relevance (MRR) model~\cite{carbonell1998use} was one of the pioneering work for promoting diversity in information retrieval tasks. In~\cite{zhang2008avoiding}, Zhang and Hurley proposed a trust-region based optimization method to maximize the diversity of recommendation list, while maintaining an acceptable level of matching quality. In~\cite{lathia2010temporal}, Lathia et al. designed and evaluated a hybrid mechanism to maximize the temporal recommendation diversity. In~\cite{zhao2012increasing}, Zhao et al. utilized the purchase interval information to increase the diversity of recommended items. 
In~\cite{qin2013promoting}, Qin and Zhu proposed an entropy regularizer to promote recommendation diversity. This entropy regularizer was incorporated in the contextual combinatorial bandit framework to diversify the online recommendation results~\cite{qin2014contextual}. Moreover, Sha et al. proposed a combinatorial optimization approach to combine the relevance of items, coverage of the user’s interests, and diversity between items for recommendation~\cite{sha2016framework}. Puthiya Parambath et al. developed a new criterion to capture both relevance and diversity for recommendation~\cite{puthiya2016coverage}. This criterion was optimized by an efficient greedy algorithm. Recently, in~\cite{antikacioglu2017post}, a maximum-weight degree-constrained
subgraph selection method was proposed to post-process the recommendations generated by collaborative filtering models, so as to increase recommendation diversity. In~\cite{cheng2017learning}, the diversified recommendation problem was formulated as supervised learning task, and a diversified collaborative filtering model was introduced to solve the optimization problems. In~\cite{chen2018fast}, a fast greedy maximum a posteriori (MAP) inference algorithm for determinantal point process was proposed to improve recommendation diversity.

\section{The Proposed Recommendation Model}
In this section, we first introduce some background about the determinantal point process model and then present the details of the proposed D$^2$RL recommendation model.
\subsection{Determinantal Point Process}
DPP is an elegant probabilistic model that has been successfully used to promote diversity in various machine learning tasks~\cite{kulesza2012determinantal}. In this work, we propose to employ DPP to promote the diversity in interactive recommendation. Specifically, for a recommendation task, a DPP on the discrete item set $\mathcal{V}=\{v_{j}\}_{j=1}^{N}$ is a probability distribution on the powerset of $\mathcal{V}$ (i.e., the set of all subsets of $\mathcal{V}$). If it assigns non-zero probability to the empty set $\emptyset$, there exists a positive semi-definite (PSD) kernel matrix $\mathbf{L} \in \mathbb{R}^{N \times N}$, such that for each subset of $\mathcal{S} \subseteq \mathcal{V}$, the probability of $\mathcal{S}$ is defined as follows:
\begin{equation}
p(\mathcal{S}) = \frac{\det(\mathbf{L}_{\mathcal{S}})}{\det(\mathbf{L} + \mathbf{I})},
\end{equation}
where $\mathbf{L}_{\mathcal{S}}$ is $\mathbf{L}$ restricted to those rows and columns which are indexed by $\mathcal{S}$, and $\mathbf{I}$ is the identity matrix.

In practice, the kernel matrix $\mathbf{L}$ can be constructed from low-rank matrices. For example, $\mathbf{L}$ can be written as a Gram matrix $\mathbf{L}=\mathbf{B}\mathbf{B}^{\top}$, where $\mathbf{B} \in \mathbb{R}^{N \times d}$ and $d \ll N$. The rows of $\mathbf{B}$ are vectors representing the properties of items. As revealed in~\cite{kulesza2012determinantal}, each row vector $\mathbf{B}_{i}$ can be empirically constructed as the product of the item relevance score $r_{i}$ and the item feature vector $\mathbf{x}_{i} \in \mathbb{R}^{1 \times d}$, i.e., $\mathbf{B}_{i} = r_{i}\mathbf{x}_{i}$. Hence, an element of the kernel matrix $\mathbf{L}$ can be written as $\mathbf{L}_{ij} = \mathbf{B}_{i}\mathbf{B}_{j}^{\top} = r_{i}r_{j} \mathbf{x}_{i} \mathbf{x}_{j}^{\top}$. Moreover, if normalization is applied to the feature vector $\mathbf{x}_{i}$, i.e., $\|\mathbf{x}_{i}\|_{2}=1$, the Cosine similarity between two items $v_{i}$ and $v_{j}$ can be calculated as $\mathbf{C}_{ij} = \mathbf{x}_{i} \mathbf{x}_{j}^{\top}$. Hence, $\mathbf{L}$ can be re-written as follows:
\begin{equation}
  \mathbf{L} = \mbox{Diag}\{\mathbf{r}\} \cdot \mathbf{C} \cdot \mbox{Diag} \{\mathbf{r}\},
\end{equation}
where $\mathbf{C} $ is the item similarity matrix, and $\mbox{Diag}\{\mathbf{r} \}$ is a diagonal matrix with the $i^{th}$ element being $r_{i}$. Following~\cite{chen2018fast}, we modify the DPP kernel matrix as follows:
\begin{equation}
  \mathbf{L} = \mbox{Diag}\{\exp(\mathbf{\alpha r})\} \cdot \mathbf{C} \cdot \mbox{Diag} \{\exp(\mathbf{\alpha r})\},
  \label{eq:dppkernel}
\end{equation}
where $\alpha=\beta / (1-\beta)$ and $\beta \in (0, 1)$. $\beta$ is a hyper-parameter used to trade-off between relevance and diversity. For simplicity, we model the relevance score of each item using a linear function of its features as follows:
\begin{equation}
    r_{i}=\mathbf{a}\mathbf{x}_{i}^{\top},
    \label{eq:relevance}
\end{equation}
where $\mathbf{a} \in \mathbb{R}^{1\times d}$ is the model parameter.

As the item similarity matrix $\mathbf{C}$ is fixed when the item feature vectors are given, according to Eq.(3) and (4), $\mathbf{a} \in \mathbb{R}^{1\times d}$ becomes the only model parameter needs to be learnt to determine the DPP kernel. Once the DPP kernel $\mathbf{L}$ is determined, different inference algorithms~\cite{kulesza2011k,chen2018fast} can be applied to find the optimal diverse while relevant item set for recommendation. Intuitively, $\mathbf{a}$ is a vector representing users' preferences. In the interactive recommendation settings, users' preferences can be dynamically learnt through the interactions between users and the recommender system. In this paper, we propose to dynamically learn $\mathbf{a}$ using a deep reinforcement learning framework.

\subsection{Deep Reinforcement Learning Framework}

In the interactive recommendation setting of D$^2$RL, the recommender system (i.e., agent) interacts with users (i.e., environment) by sequentially recommending a set of diverse while relevant items to maximize its cumulative rewards in the long run. The interactive recommendation process can be modeled as a Markov Decision Process (MDP). We define the key components of the MDP of D$^2$RL as follows:
\begin{itemize}

    \item \emph{State Space}: A state $\mathbf{s}_{t}$ is determined by the recent $\ell$ items that the user has interacted before time step $t$ and the user's personal features;

    \item \emph{Action Space}: An action $\mathbf{a}_{t}$ is the parameter used to construct the DPP kernel matrix, as shown in Eq.~\eqref{eq:relevance}, which captures the user's dynamic preferences;

    \item \emph{Reward}: Once the recommender system chooses an action $\mathbf{a}_{t}$ based on its current state $\mathbf{s}_{t}$, a set of diverse while relevant items can be sampled from the determined DPP kernel and recommend to the user. Then, the user would provide her feedbacks on the recommended items, e.g., click or not click on the items. The recommender system receives these feedbacks as the immediate reward $r(\mathbf{a}_{t}, \mathbf{s}_{t})$ to update its recommendation policy;

    \item \emph{Discount factor}: The discount factor $\gamma \in [0, 1]$ is used to measure the present value of future rewards. When $\gamma$ is set to 0, the recommender system only considers the immediate rewards but ignores the future rewards. On the other hand, when $\gamma$ is set to 1, the recommender system counts all the future rewards as well as the immediate rewards.
\end{itemize}

\begin{figure}
    \centering
    \includegraphics[width=3.3in]{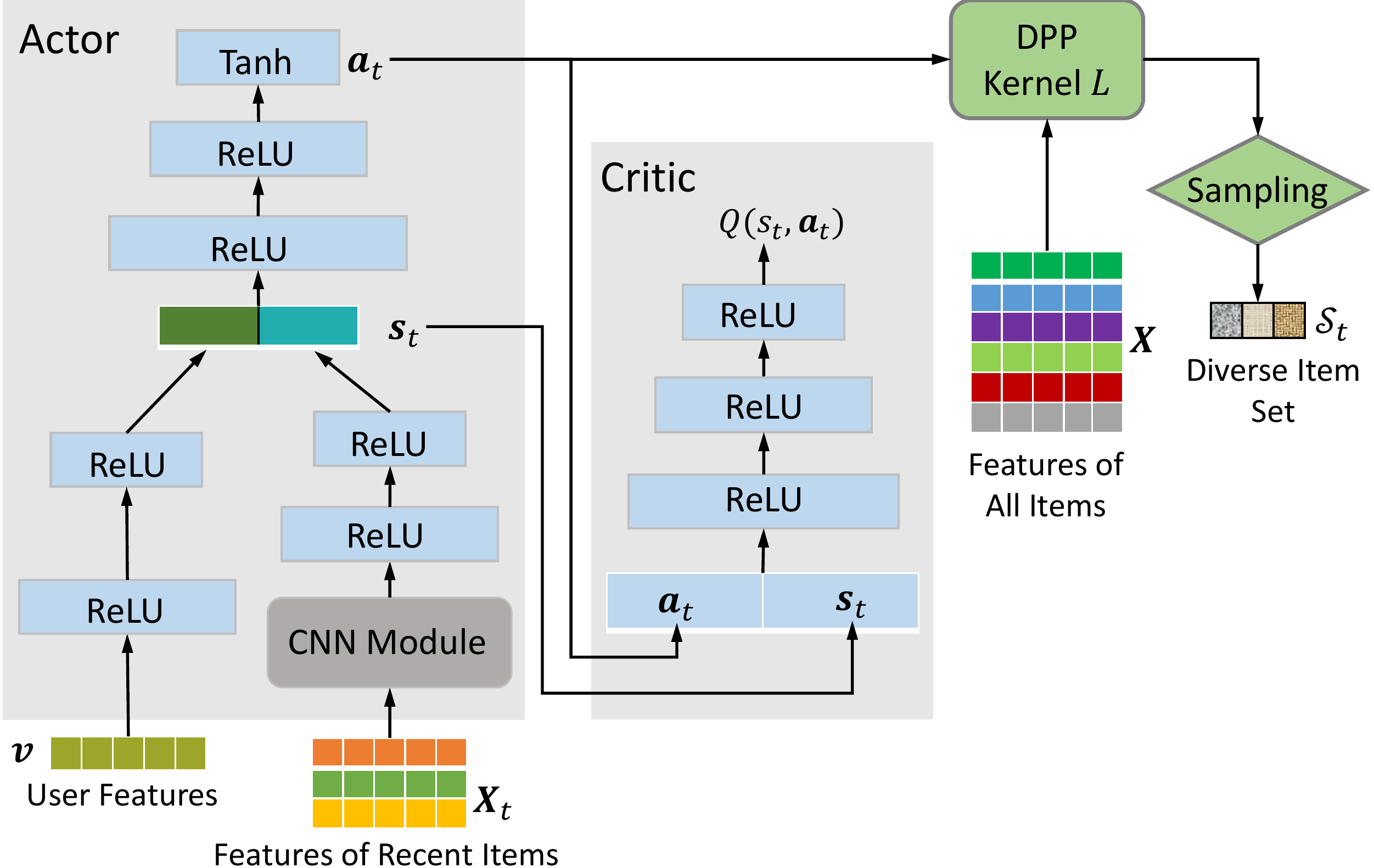}
    \vspace{-5px}
    \caption{The proposed deep reinforcement learning framework.}
    \vspace{-8px}
    \label{fig:framework}
\end{figure}

The proposed D$^2$RL model is built based on the Actor-Critic reinforcement learning framework~\cite{sutton2018reinforcement}. Figure~\ref{fig:framework} shows the overall structure of D$^2$RL. Next, we detail the actor and the critic components of D$^2$RL.

\begin{itemize}

  \item \textbf{\emph{Actor Network}}: Given the current state $\mathbf{s}_{t}$, the actor network employs deep neural networks to generate an action $\mathbf{a}_{t}$. As shown in Figure~\ref{fig:framework}, the state $s_{t}$ is determined by the user's personal features and her recent interests.
      Specifically, the user's personal features are fed into two fully-connected ReLU neural network layers to get more abstract user features $\mathbf{v}$. Meanwhile, at time step $t$, we collect the features of $\ell$ recent items that the user has interacted before $t$, and stack them into a feature matrix $\mathbf{X}_{t} \in \mathbb{R}^{\ell \times d}$. Intuitively, we treat the matrix $\mathbf{X}_{t}$ as an ``image" of the feature space, and use CNN (consisting of two convolutional layers) to capture the sequential patterns of the user's interests by extracting the local features of the ``image"~\cite{tang2018personalized}. The output of the CNN module is then fed into two fully-connected ReLU neural network layers to get higher level features $\mathbf{z}_{t}$. Then, we concatenate $\mathbf{z}_{t}$ with user features $\mathbf{v}$ and define the \emph{state representation} as $\mathbf{s}_{t} = [\mathbf{v}, \mathbf{z}_{t}]$.

      After that, $\mathbf{s}_{t}$ is fed into several fully connected layers to generate the action $\mathbf{a}_{t}$. In the first two neural network layers, we use ReLU as the activation function, and we use Tanh activation function in the output layer to make $\mathbf{a}_{t}$ bounded. As we are interested in optimizing the DDP kernel to generate diverse recommendations, the action representation $\mathbf{a}_{t}$ is essentially the parameter used to build the DPP kernel $\mathbf{L}$ at time $t$. Once $\mathbf{L}$ is determined, we utilize the fast greedy MAP inference algorithm~\cite{chen2018large} to do sequentially sampling of a set of diverse while relevant items as recommendations to the user.

  \item \textbf{\emph{Critic Network}}: The critic network is designed to approximate the Q-value function $Q^{\pi}(\mathbf{s}_{t}, \mathbf{a}_{t})$, which estimates the quality of the learnt DPP-based recommendation policy $\pi$. As shown in Figure~\ref{fig:framework}, the input of the critic network is the action representation $\mathbf{a}_{t}$ and the state representation $\mathbf{s}_{t}$ that considers both the user profile and the user's recent interests. $\mathbf{a}_{t}$ and $\mathbf{s}_{t}$ are then fed into several fully connected ReLU neural network layers to predict the Q-value. We denote the critic network by $ \varphi(\mathbf{s}_{t}, \mathbf{a}_{t})$ and the network parameters by $\theta_{\varphi}$.

\end{itemize}

\begin{algorithm}[t]
  \caption{DDPG algorithm for D$^{2}$RL}\label{algorithm:DDPG}
  \begin{algorithmic}[1]
    \STATE Randomly initialize the actor network $\phi(\cdot)$ and critic network $\varphi(\cdot)$ with the weights $\theta_{\phi}$ and $\theta_{\varphi}$;
    \STATE Initialize the target networks $\tilde{\phi}(\cdot)$ and $\tilde{\varphi}(\cdot)$ with weights $\tilde{\theta}_{\phi} \leftarrow \theta_{\phi}$ and $\tilde{\theta}_{\varphi} \leftarrow \theta_{\varphi}$;
    \STATE Initialize the replay buffer;
    \FOR{$episode=1, 2, \dots$}
        \STATE Receive an initial observed state $\mathbf{s}_{1}$;
        \FOR{$t=1, 2, \dots, T$}
            \STATE Derive action $\mathbf{a}_{t}$ from actor network $\phi(\cdot)$;
            \STATE Apply exploration policy to $\mathbf{a}_{t}$: $\mathbf{a}_{t}\leftarrow \mathbf{a}_{t}+\mathcal{N}_{t}$;
            \STATE Construct DPP kernel matrix $\mathbf{L}$ using Eq.(3), (4);\\
            \STATE Find the diverse item set $\mathcal{S}_{t}$ using the fast greedy MAP inference algorithm;\\
            \STATE Compute reward $r_{i}$ and observe new state $s_{t+1}$;\\
            \STATE Store transition sample $(\mathbf{s}_{t}, \mathbf{a}_{t}, r_{t}, \mathbf{s}_{t+1})$ to replay buffer;\\
            \STATE Sample a mini-batch of $N$ transitions from replay buffer with replay sampling technique;\\
            \STATE Update the critic network $\varphi(\cdot)$ by minimizing the loss in Eq.~\eqref{eq:criticloss};\\
            \STATE Update the weights $\theta_{\phi}$ of actor network $\phi(\cdot)$ using the sampled gradient in Eq.~\eqref{eq:actorUpdate};\\
            \STATE Update weights of target networks using Eq.~\eqref{eq:softupdate};
        \ENDFOR
    \ENDFOR
  \end{algorithmic}
\end{algorithm}

The Deep Deterministic Policy Gradient (DDPG)~\cite{lillicrap2015continuous} algorithm is used to train the D$^2$RL framework. We also employ the target network technique in D$^2$RL. In DDPG, the critic network is learned using the Bellman equation in Q-learning. The network parameter $\theta_{\varphi}$ is updated by minimizing the following loss function:
\begin{align}
    \label{eq:criticloss}
    \ell(\theta_{\varphi})=\frac{1}{N}\sum_{i=1}^{N}\big( y_{i} - \varphi(\mathbf{s}_{i}, \mathbf{a}_{i}) \big)^{2},
\end{align}
where $y_{i} = r_{i} + \gamma \tilde{\varphi}(\mathbf{s}_{i+1}, \tilde{\phi}(\mathbf{s}_{i+1}))$, $N$ is the size of the sampled mini-batch, $\tilde{\phi}$ and $\tilde{\varphi}$ are the target actor network and target critic network, respectively. The actor network is updated by using sampled policy gradient as follows:
\begin{align}
\label{eq:actorUpdate}
\nabla_{\theta_{\phi}} \phi|s_{i} = \frac{1}{N}\sum_{i=1}^{N}\nabla_{\mathbf{a}} \varphi(\mathbf{s}, \mathbf{a})|_{\mathbf{s}=\mathbf{s}_{i}, \mathbf{a}= \phi(s_{i})} \nabla_{\theta_{\phi}} \phi(s)|_{s=s_{i}}.
\end{align}
Moreover, the network parameters of the target networks $\tilde{\theta}_{\phi}$ and $\tilde{\theta}_{\varphi}$ are updated using the soft updating rule as follows:
\begin{align}
\tilde{\theta}_{\phi} \leftarrow \tau \theta_{\phi} + (1-\tau) \tilde{\theta}_{\phi}, \nonumber\\
\tilde{\theta}_{\varphi} \leftarrow \tau \theta_{\varphi} + (1-\tau) \tilde{\theta}_{\varphi}
\label{eq:softupdate}
\end{align}
The details of the DDPG based learning algorithm are summarized in Algorithm~\ref{algorithm:DDPG}.

\section{Experiments}
In this section, we first evaluate the performance of the proposed D$^2$RL model by performing offline experiments on two real world datasets. Then, we conduct simulations to evaluate the online performance of D$^2$RL.

\subsection{Experimental Setting}
\subsubsection{Datasets}
The experiments are conducted on two public datasets: MovieLens-100K and MovieLens-1M~\footnote{https://grouplens.org/datasets/movielens/}. MovieLens-100K consists of 100,000 ratings given by 943 users to 1,682 movies. MovieLens-1M contains 1,000,209 ratings given by 6,040 users to 3,706 movies. There are 18 movie categories in both datasets, and each movie may belong to more than one categories. In each dataset, we kept the ratings larger than 3 as positive feedback. We sort all the positive feedback in time order, and use the former 80\% positive feedback for model training and the remaining 20\% positive feedback for model testing. Moreover, we also remove the users that have less than 10 positive feedbacks in the training data. Table~\ref{tab:dataset} summarizes the statistics of the experimental datasets.
\begin{table}[h!]
    \centering
    \small
    \caption{Statistics of the experimental datasets.}
   \vspace{-8px}
    \label{tab:dataset}
    \begin{tabular}{|l|c|c|c|c} \hline
    Dataset & \# users & \# items & \# Interactions \\\hline
    Movielens-100K & 716 & 1,374 & 45,447 \\ \hline
    Movielens-1M & 5,218 & 3,467 & 510,940  \\ \hline
    \end{tabular}
    \vspace{-8px}
\end{table}

\subsubsection{Evaluated Algorithms}
We compare the proposed D$^2$RL method with the following recommendation models: (1) \textbf{\emph{BPRMF}}~\cite{rendle2009bpr}: This is a matrix factorization method that uses pairwise learning to optimize the recommendation accuracy of top-$N$ items. (2) \textbf{\emph{LMF}}~\cite{johnson2014logistic}: This is a logistic matrix factorization recommendation approach that exploits users' implicit feedback by modeling the user-item interaction probability. (3) \textbf{\emph{Caser}}~\cite{tang2018personalized}: This is a deep learning based sequential recommendation method which uses CNN to extract the sequential patterns of users' behaviors. (4) \textbf{\emph{C$^2$UCB}}~\cite{qin2014contextual}: This is a contextual bandit based interactive recommendation method, which promotes the recommendation diversity by employing an entropy regularizer.

For BPRMF and LMF, we empirically set the dimensionality of latent space to 30, considering both the recommendation accuracy and efficiency. The regularization parameters are chosen from $\{10^{-4}, 10^{-3}, \cdots, 10^{2}\}$, and the optimal learning rates are chosen from $\{2^{-5}, 2^{-4}, \cdots, 2^{2}\}$. For Caser, we set the dimensionality of the embedding layer to 30, the Markov order $L$ to 5, and the target number $T$ to 2. In addition, we use the latent features of users and items learnt by BPRMF as the user and item features in the interactive recommendation methods, i.e., C$^2$UCB and D$^2$RL. In C$^2$UCB, the regularization parameter $\lambda$ of the entropy regularizer is selected from $\{0.01, 0.1, 0.5, 1.0, 10.0\}$. For D$^2$RL, we set the number of recent items $\ell$ in each state to 5. The parameter $\beta$ that balances relevance and diversity in the DPP kernel matrix is chosen from $\{0.1, 0.2, \cdots, 0.9\}$. We utilize Adam to optimize the actor and critic networks, and set the learning rate to 0.001. In addition, we set the discount factor $\gamma$ of D$^2$RL to 0.95.

\begin{algorithm}[t]
  \caption{Offline Evaluation for D$^2$RL}\label{algorithm:offlineEval}
  \begin{algorithmic}[1]
    \STATE Initialize the actor network by minimizing the distance between its output and the embedding of the next interacted item;
    \STATE Initialize the critic network by fixing the actor network;
    \STATE Train the actor-critic network following DDPG until converge;
    \STATE Initialize the user's state $\mathbf{s}_{0}$ with the last $\ell$ interacted items in the training data:
    \STATE Initialize the list of displayed items $\mathcal{B}$ with the last $\ell$ interacted items in the training data;
    \STATE Collect candidate items $\mathcal{V}_{0}$ by excluding the user's interacted items the training data;
    \STATE Load the user's interacted items $\mathcal{V}_{test}$ in testing data;
    \FOR{$t=1, 2, \dots, T$}
        \STATE Generate action by the actor network $\mathbf{a}_{t} \leftarrow \phi(\mathbf{s}_{t})$;
        \STATE Construct DPP kernel matrix $\mathbf{L}$ and find the diverse item set $\mathcal{S}_{t}$ by the fast greedy inference algorithm;\\
        \FOR{$v_{i} \in \mathcal{S}_{t}$}
            \IF{$v_{i} \in \mathcal{V}_{test}$}
                \STATE Remove the first item in $\mathcal{B}$, and append $v_{i}$ to $\mathcal{B}$;
            \ENDIF
        \ENDFOR
        \STATE Compute recommendation performances using Eq. (8) and (9);
        \STATE Generate new state $\mathbf{s}_{t+1}$ with $\mathcal{B}$;
        \STATE $\mathcal{V}_{0} \leftarrow \mathcal{V}_{0} \setminus \mathcal{S}_{t}$;
        \ENDFOR
  \end{algorithmic}
\end{algorithm}

\subsubsection{Metrics}
The performance of recommendation algorithms is evaluated from two aspects, i.e., accuracy and diversity. The recommendation accuracy at each recommendation epoch is defined as follows:
\begin{equation}
    \mbox{Precision}(t) = \frac{|\mathcal{S}_{t} \cap \mathcal{V}_{test}|}{|\mathcal{S}_{t}|},
\end{equation}
where $\mathcal{S}_{t}$ is the set of items recommended to the user at epoch $t$, $|\cdot|$ is the cardinality of a set, and $\mathcal{V}_{test}$ denotes the set of items the user has interacted with in the testing data. The diversity of the recommendation results at epoch $t$ is measured by the intra-list distance (ILD)~\cite{zhang2008avoiding} metric, which is defined as follows:
\begin{equation}
    \mbox{Diversity}(t)=1 - \frac{2}{|\mathcal{S}_{t}|(|\mathcal{S}_{t}| - 1)}\sum_{v_{i} \in \mathcal{S}_{t}} \sum_{v_{j} \in \mathcal{S}_{t}, i \neq j} s_{ij},
\end{equation}
where $s_{ij}$ denotes the similarity between two items $v_{i}$ and $v_{j}$. For fair comparison, we define $s_{ij}$ by the Jaccard similarity between the categories of two items $v_{i}$ and $v_{j}$. In the experiments, we empirically set the size of $\mathcal{S}_{t}$ to 5.

\subsubsection{Setup}
We adopt two different settings to evaluate the recommendation algorithms: offline evaluation on public datasets and online evaluation with a simulator.

\begin{algorithm}[t]
  \caption{Simulator for Online User Behavior}\label{algorithm:onlineEval}
  \begin{algorithmic}[1]
    \STATE Observe all the previously interacted items $\mathcal{R}$ of the user;
     \FOR{$v_{i} \in \mathcal{S}_{t}$}
         \STATE Compute the user's interaction probability $p_{i}$ with $v_{i}$ using pre-trained LMF;
         \IF{$|\mathcal{R}| > 0$}
             \STATE $p_{i}\leftarrow \delta p_{i} + (1-\delta)\frac{1}{|\mathcal{R}|}\sum_{v_{j} \in \mathcal{R}} (1 - \mathbf{C}_{ij}) $; \
         \ENDIF
            \IF{$p_{i} > \rho$}
                \STATE Set the reward $r_{i} \leftarrow 1$, $\mathcal{R} \leftarrow \mathcal{R} \cup \{v_{i}\}$ ;
            \ELSE
                \STATE Set the reward $r_{i} \leftarrow 0$;
            \ENDIF
        \ENDFOR
  \end{algorithmic}
\end{algorithm}

In offline evaluation, we compare D$^2$RL with all other baseline mehods. For each non-interactive recommendation approach (i.e., BPRMF, LMF, and Caser), we first learn the recommendation model using the training data, and then rank all candidate items by the learned model. At each recommendation epoch, we choose a set of top ranked items $\mathcal{S}_{t}$ from the ranked candidates as recommendations, and these items will be removed from the candidate set once they have been recommended. The recommendation accuracy will be measured by comparing recommended items with the ground truth in the testing data. Note that the non-interactive recommendation methods would not update the learned models in the recommendation process. However, in C$^2$UCB, the bandit parameter is updated by using the user's feedback on the recommendation results. For the offline training of D$^2$RL, we first initialize the actor network by minimizing a L2-loss between its output and the embedding of the next interacted item by randomly sampling state-action transition samples from the replay buffer. After that, we fix the actor network for some time and update the critic network alone. When both the actor network and the critic network are well initialized, they will follow DDPG (Algorithm~\ref{algorithm:DDPG}) to update iteratively. Once the two networks converge, we fix the learned network parameters in the recommendation process. The details of the offline evaluation process for D$^2$RL are summarized in Algorithm~\ref{algorithm:offlineEval}.

For online evaluation, we build a simulator to compare D$^2$RL with C$^2$UCB. The online interactive recommendation process of D$^2$RL follows Algorithm~\ref{algorithm:DDPG} and the feedback on recommended items is generated by a simulator as summarized in Algorithm~\ref{algorithm:onlineEval}. In this simulator, we first use the pre-trained LMF~\cite{johnson2014logistic} model on training data to predict user preference on the recommended items. Then, we use MMR~\cite{carbonell1998use} to simulate the user's feedback probability by considering both the user preference and the diversity of recommended items. The MMR parameter $\delta$ simulates a person's personal consideration between relevance and diversity, and a higher $\delta$ simulates a person who prefers more relevant and less diverse recommendations. The parameter $\rho$ is a decision threshold, and positive feedback will be provided if the user's feedback probability is greater than $\rho$. We empirically set the MMR parameter $\delta$ for each user randomly in the range $(0, 1)$ and the threshold $\rho$ to 0.5. A good interactive recommender system should be able to learn from the simulated user behaviours and make accurate while diverse recommendations. For online evaluation, the accuracy is evaluated by comparing the recommended results with simulated feedbacks.

\begin{figure}[t!]
    \centering
   \includegraphics[width=3.25in]{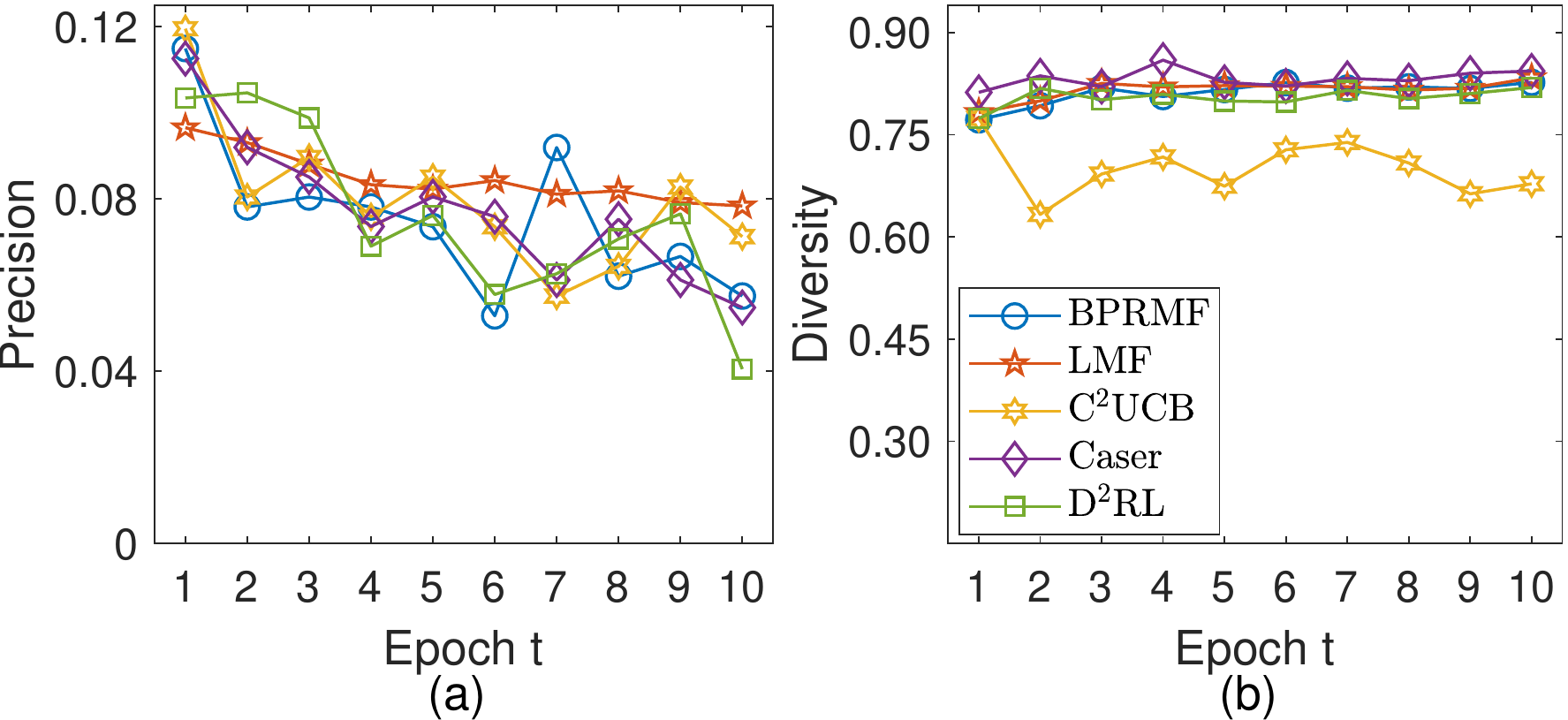}
   \label{fig:ml1mprec}
   \vspace{-8pt}
    \caption{Offline evaluation results on Movielens-100K.}
    \vspace{-8px}
    \label{fig:ml100kOffline}
\end{figure}
\begin{figure}[t!]
    \centering
   \includegraphics[width=3.25in]{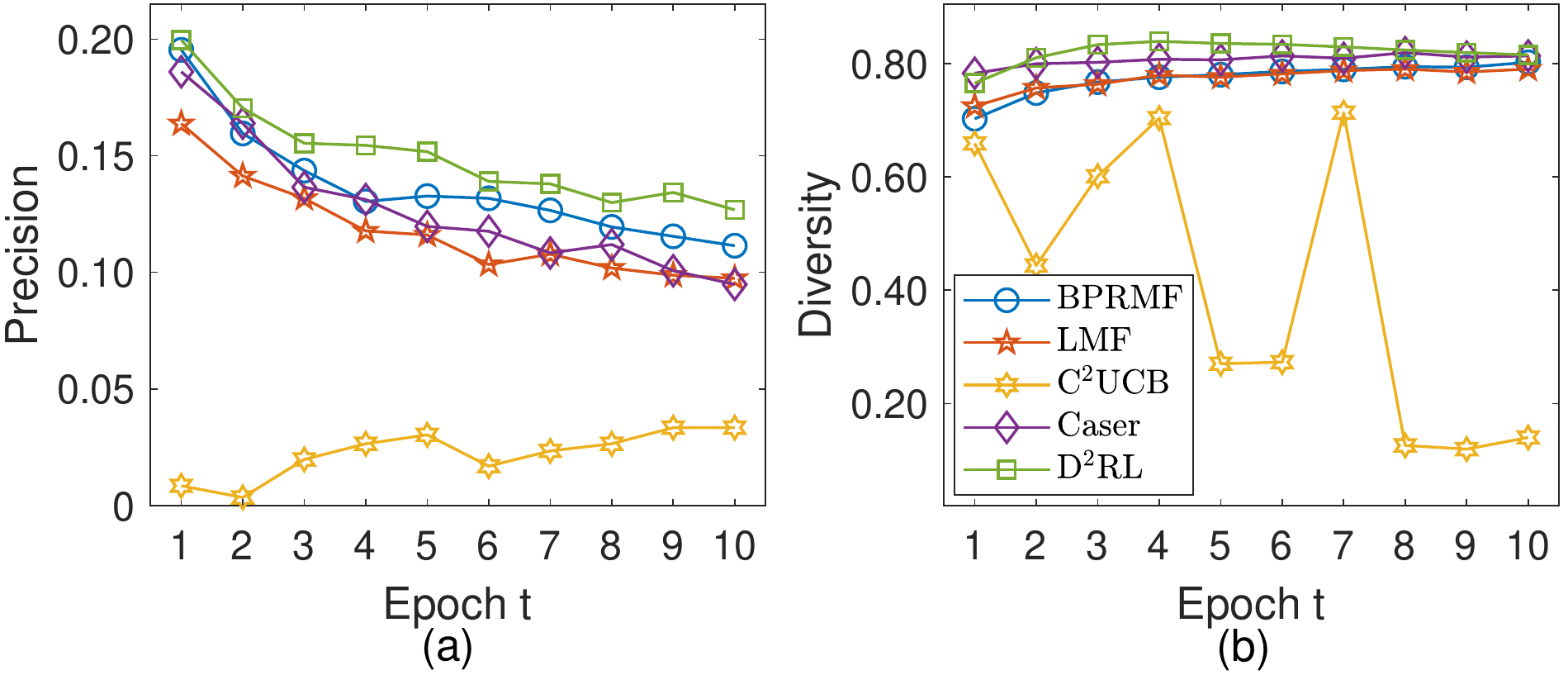}
   \label{fig:ml1mprec}
   \vspace{-8pt}
    \caption{Offline evaluation results on Movielens-1M.}
    \vspace{-8px}
    \label{fig:ml1mOffline}
\end{figure}

\subsection{Experimental Results}
\subsubsection{Offline Performance}
The offline performance of the evaluated recommendation algorithms on Movielens-100K and Movilens-1M datasets are summarized in Figure~\ref{fig:ml100kOffline} and Figure~\ref{fig:ml1mOffline}, respectively. It can be observed from Figure~\ref{fig:ml1mOffline}(a) that D$^2$RL consistently outperforms all the baselines in terms of recommendation accuracy at all recommendation epochs on Movielens-1M dataset. Moreover, from Figure~\ref{fig:ml1mOffline}(b), we can also note that D$^2$RL also consistently generates more diverse recommendations than the baselines at all recommendation epoches.

However, the performance of D$^2$RL is not as superior on Movielens-100K dataset. As shown in Figure~\ref{fig:ml100kOffline}(a), D$^2$RL sometimes achieves poorer recommendation accuracy at certain recommendation epoches comparing with other baselines. One potential reason is that the critic network is not able to accurately estimate the quality of the learned recommendation policy. From Figure~\ref{fig:ml100kOffline}(b), we can note that D$^2$RL achieves comparable performance with the baselines in terms of recommendation diversity. In the offline setting, the user's feedback on the recommendation results is fixed, which limits the effectiveness of interactive recommender systems in capturing users' dynamic preferences on items. To remedy this problem, we conduct simulations of online experiments to study the effectiveness of D$^2$RL, based on Movielens-100K and Movielens-1M datasets.

\subsubsection{Simulated Online Performance}
\begin{figure}[t!]
    \centering
   \includegraphics[width=3.3in]{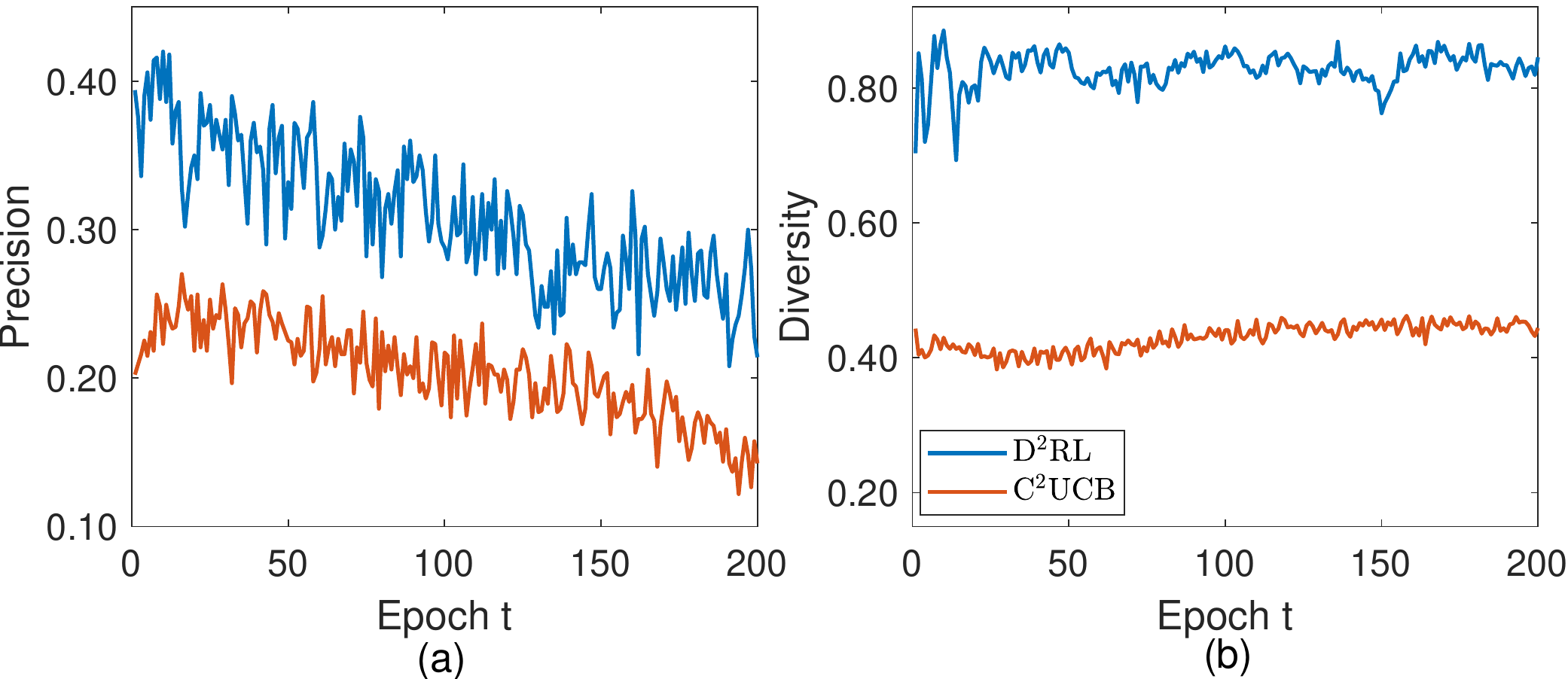}
   \label{fig:ml1mprec}
   \vspace{-4pt}
    \caption{Online simulation performances on Movielens-100K.}
    \vspace{-6px}
    \label{fig:ml100kOnline}
\end{figure}

\begin{figure}[t!]
    \centering
   \includegraphics[width=3.3in]{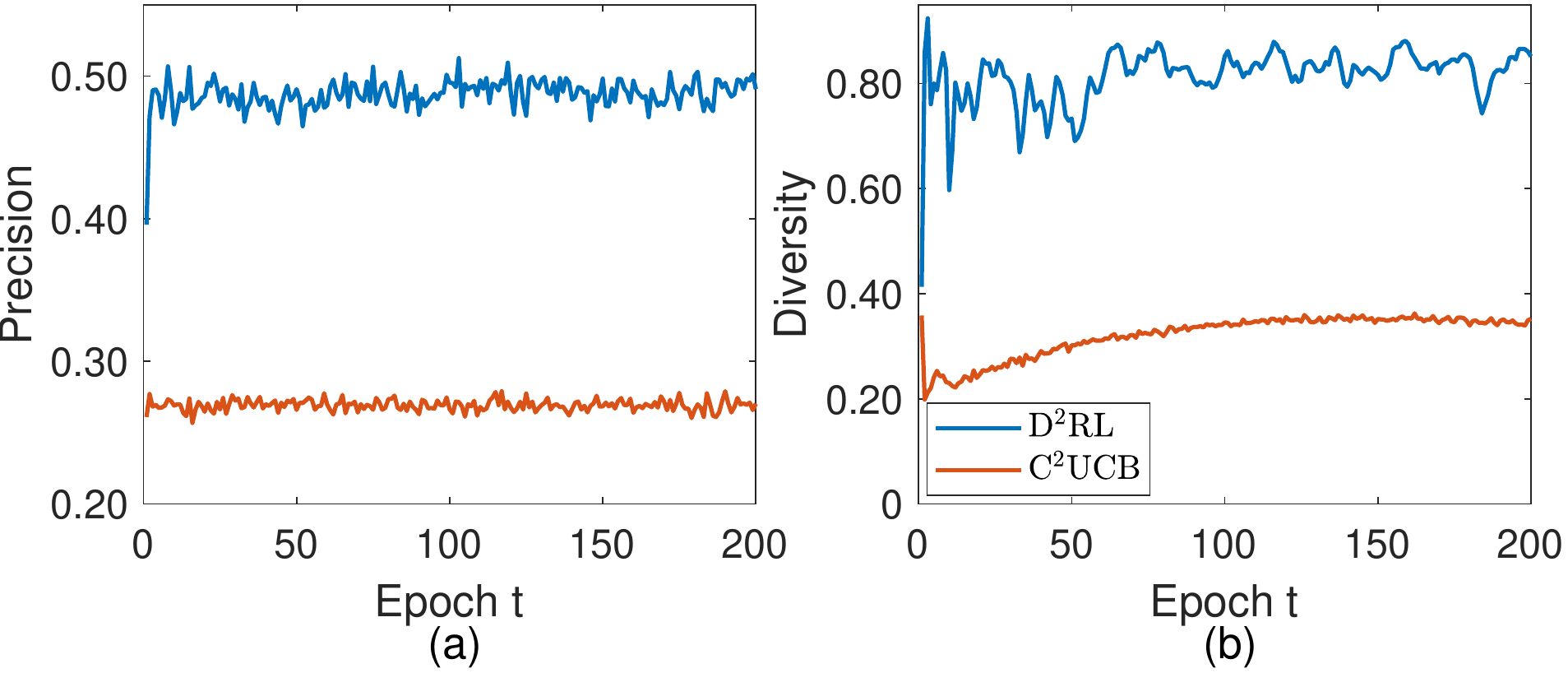}
   \label{fig:ml1mprec}
   \vspace{-9pt}
    \caption{Online simulation performances on Movielens-1M.}
    \vspace{-10px}
    \label{fig:ml1mOnline}
\end{figure}

The simulated online performance of D$^2$RL and C$^2$UCB on Movielens-100K and Movilens-1M datasets are summarized in Figure~\ref{fig:ml100kOnline} and Figure~\ref{fig:ml1mOnline}, respectively. As shown in Figure~\ref{fig:ml100kOnline}(a), the online recommendation accuracy of D$^2$RL and C$^2$UCB both fluctuate at all recommendation epochs, but D$^2$RL consistently achieves better recommendation accuracy than C$^2$UCB at all recommendation epochs. Moreover, from Figure~\ref{fig:ml100kOnline}(b), we can also observe that the online recommendation results generated by D$^2$RL are much more diverse than those generated by C$^2$UCB. On average, D$^2$RL outperforms C$^2$UCB by 52.13\% and 92.79\%, in terms of Precision and Diversity, over all recommendation epoches in the simulated experiments on Movielens-100K datasets.

Superior performance of D$^2$RL can also be observed on Movielens-1M dataset. As shown in Figure~\ref{fig:ml1mOnline}(a), both D$^2$RL and C$^2$UCB achieve stable performance on recommendation accuracy, and D$^2$RL significantly outperforms C$^2$UCB consistently at all recommendation epochs. Moreover, Figure~\ref{fig:ml1mOnline}(b) shows that the recommendation diversity of D$^2$RL fluctuates at first and then becomes steady quickly, which is significantly higher than that of C$^2$UCB. On average, D$^2$RL outperforms C$^2$UCB by 76.04\% and 156.58\% in terms of Precision and Diversity, over all recommendation epoches in the simulation experiments on Movielens-1M dataset.

\section{Conclusion}
In this paper, we have proposed a novel interactive recommendation model, namely Diversity-promoting Deep Reinforcement Learning (D$^2$RL). Specifically, D$^2$RL uses actor-critic reinforcement learning framework to model the interactions between users and the recommender system. Moreover, the determinantal point process (DPP) model has been incorporated in the recommendation process to promote the diversity of recommendation results. Both empirical experiments on real world datasets and simulated online experiments have been performed to demonstrate the effectiveness of D$^2$RL. The future work will focus on the following directions. Firstly, we would like to directly learn the DPP kernel matrix from the data. Secondly, we are also interested in developing more efficient DPP inference algorithms for generating relevant yet diverse recommendations.

\bibliographystyle{named}
\bibliography{ijcai19}

\end{document}